\documentclass[runningheads]{llncs}
\usepackage{graphicx}
\usepackage{pgfplots}
\begin{document}
\title{Effect of intervention policies in an agent based spatial epidemic model of Gwalior\thanks{This work was carried out at ABV- IIITM Gwalior, India with funding from SERB, DST, Govt. of India}}
%
%
\author{W. Wilfred Godfrey\inst{}\orcidID{0000-0002-0720-2647}}
\authorrunning{W. Wilfred Godfrey}
%
\institute{Atal Bihari Vajpayee Indian Institute of Information Technology and Management, Gwalior, MadhyaPradesh, India 
\email{godfrey@iiitm.ac.in}}
\maketitle              
\begin{abstract}
Covid-19 has ravaged the entire world and it may not be the last such to ravage the world. COMOKIT \cite{comok} is an agent based spatial modeling tool to study the effect of covid -19 in a geographical area by creating heterogenous synthetic agents and their behaviours. This paper presents comokit based case study on Gwalior region with respect to various intervention policies to curb the spread of the disease.

\keywords{COMOKIT, Multi-Agent Systems  \and Spatial Modeling \and Geographical Information System}
\end{abstract}
\section{Introduction}
\subsection{Background}

What is the effect on the pandemic in a semi urban area such as Gwalior with both the rural and urban population. How does the disease dynamics vary with different intervention policies in the context of Gwalior region. As the disease dynamics is explored, the effect of disease regulatory policies on the covid spread is also observed and checked.

A spatial approach towards measures on containing the disease spread is very useful since it is not possible to microexamine the spatial effect from the data that is available in the public domain. The other factors that have an effect in the disease spread are the social, cultural, polical in nature which are also difficult to mathematically model and assess. Within this context, agent based modeling approach is a feasible option to incorporate the many details of real world into its simulation model and perform the necessary experimentations.

There have been a large number of models that have been proposed in the recent times varying from systems to study the macro effect of the pandemic to the systems to study micro effect of the pandemic. Agent based system help study the micro effect of the pandemic by modeling individual human entities, their simple and compound behaviours so as to construct and study the complex phenomenon that evolves.

COMOKIT (COVID-19 Modelling Kit) \cite{gaudou2020comokit} is an agent based  near-realistic spatial framework. The major submodels included in the COMOKIT are
\begin{enumerate}
    \item Person to person transmission sub-model
    \item Environmental transmission sub-models
    \item Human mobility sub-model
    \item Policy intervention sub-model
\end{enumerate}

\subsection{Objectives}
The objectives of this work are as follows:
\begin{enumerate}
    \item To present the results on applying the agent based COMOKIT model on the various rural to urban regions in and around around Gwalior
    \item To present the results on applying the agent based COMOKIT model on the various rural to urban regions in and around Gwalior with various covid restrictions.
\end{enumerate}

The rest of the work is organized as follows. Section 2 presents a literature survey with the literature gap in this direction. Section 3 presents some of the statistics and details with respect to the covid-related scenario in Gwalior. Section 4 presents the  COMOKIT model as well as the results obtained by applying the COMOKIT model on the gis maps of Gwalior followed by concluding remarks in section 5.

\section{Related Work}
This section discusses about COMOKIT model as well as various works carried out in literature using COMOKIT model.
\subsection{COMOKIT (COVID-19 Modelling Kit)}
COMOKIT (COVID-19 Modelling Kit) \cite{gaudou2020comokit} is an agent based model developed to aid the Government of Vietnam in making intervention based decisions with respect to small towns of around 10000 inhabitants. The model comprises of spatially dependent, interconnected and individual agents with private attributes such as age, gender, employment and disease status. Each agent is fit with an agenda which is a set of activity which is dependent on its private attributes. Buildings are special type of entities, includes hospitals to contain sick and critical diseased agents.  

An agent based simulation is executed over several "steps" with each step representing 1 hour. Contagious agents with conditional probability infect susceptible ones within same building. Special agents called Authority agents test inhabitants for infection.

\subsection{Literature related to COMOKIT}
Gaudou et al \cite{gaudou2020comokit,drogoul2020designing} have presented the COMOKIT model and also performed experiments to explore the potential of the model. Brugiere et al. \cite{brugiere2010odd} have presented an O.D.D. description of the COMOKIT model. The ‘ODD’ (Overview, Design concepts, and Details) protocol standardizes descriptions of individual-based and agent-based models. Chapuis et al.\cite{chapuis2021comokit} have studied the effect of COMOKIT with respect to non-pharmaceutical interventions for reducing the spread of covid-19 with cases such as COMOKIT Azur and COMOKIT Camps on the transmission of covid in
refugee camps.
Chapuis et al.\cite{chapuis2021using} further studied the effect of morpho-functional organization of cities on the spread of COVID-19. COMOKIT Albatross \cite{kim2021comokit} was one of the first country-scale application of COMOKIT in the Netherlands using data for the construction of individuals’ agendas accounting for covid variants, vaccination intervention, better dashboard for visualizing data etc.

\subsection{ Literature related to Agent based systems for India }
Narassima et al.\cite{narassima2020agent1} have modeled the COVID-19 transmission dynamics using an agent based approach for Telangana and India \cite{narassima2020agent1}, \cite{narassima2021agent} using synthetic population.  Klein et al.\cite{klein2020covid} have provided a projection of estimates of infection for India based on IndSim, an agent based system. Shubhada et al. \cite{agrawal2020city} used an agent based simulator for studying effect of covid-19 pandemic in Bengaluru and Mumbai. Suryavanshi et al. \cite{suryawanshi2021city} have used an agent based approach to study the effect of the disease in a generic city of population around 10000 parameterizing their model on Kolkata city. Bharat et al. \cite{barat2021agent} have examined the effect of interventions with the characteristics of the city of  Pune, Maharashtra using an agent based model. 

Among these only \cite{suryawanshi2021city} and \cite{agrawal2020city} are spatio-temporal models.

\section{Methodology}

\begin{figure}
  \begin{tabular}{ r c c }
    \phantom{Dataset~1} &
\includegraphics[width=0.5\textwidth]{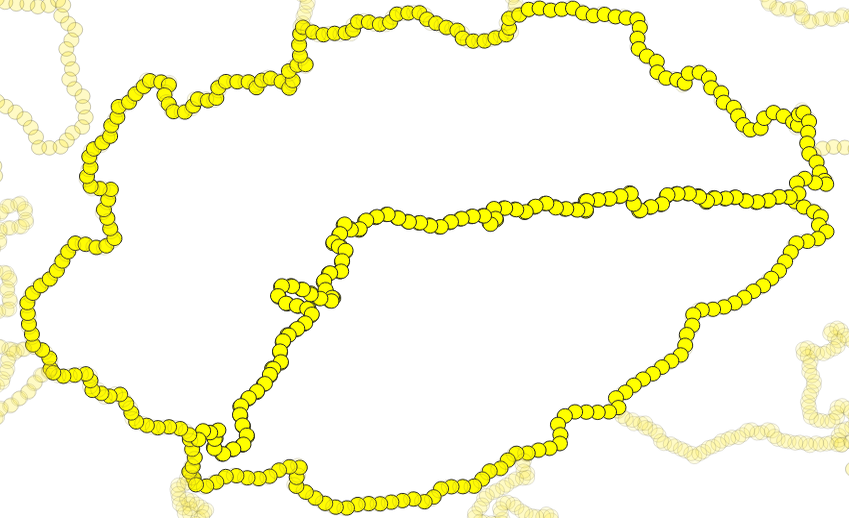} &
\includegraphics[width=0.5\textwidth]{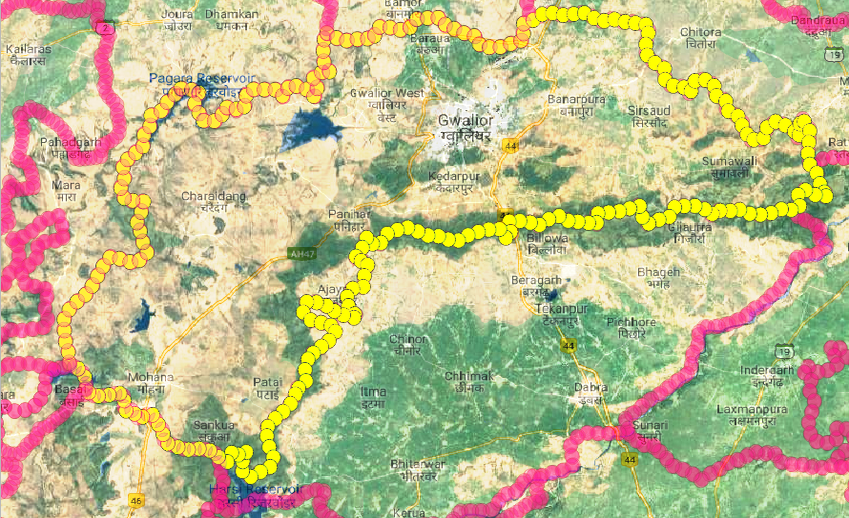} 
\\ \\
    & (a)Gwalior District &
      (b)Northern Gwalior
  \end{tabular}
  \caption{The Gwalior District and the Northern part of Gwalior district comprising Ghatigaon and Morar blocks wihin the yellow contour region.} \label{figure1}
\end{figure}

\begin{figure}
  \begin{tabular}{ r c c }
  \phantom{Dataset~2} &
      \includegraphics[width=0.5\textwidth]{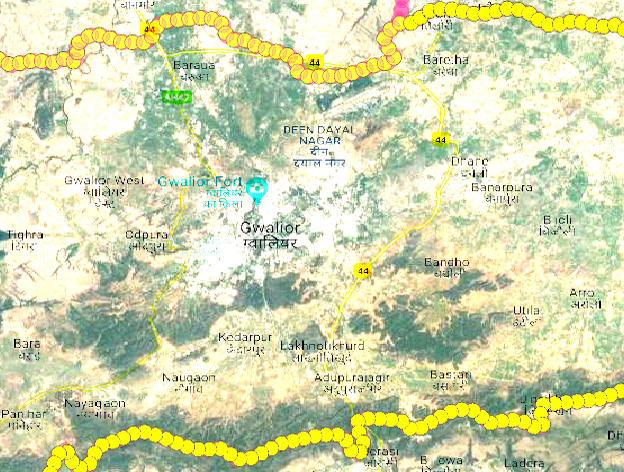} &
      \includegraphics[width=0.5\textwidth]{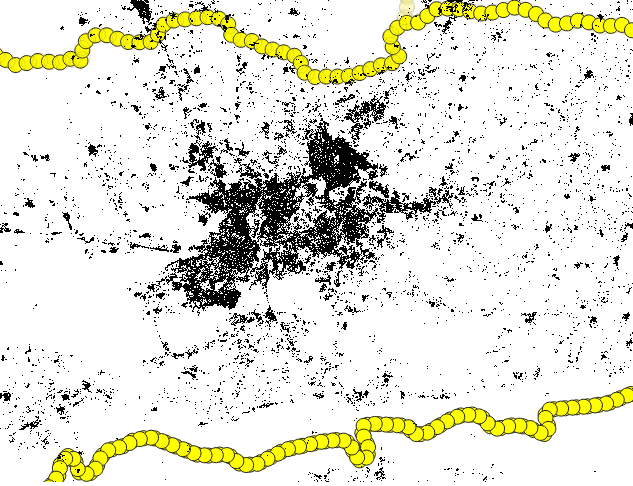}
\\ \\
    & (a)Gwalior City Locations Map &
      (b)Gwalior City Population Map \\  \\
    \phantom{Dataset~1} &
\includegraphics[width=0.5\textwidth]{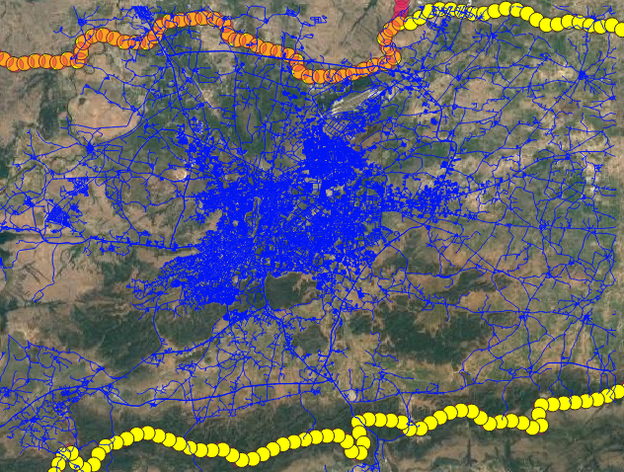} &
\includegraphics[width=0.5\textwidth]{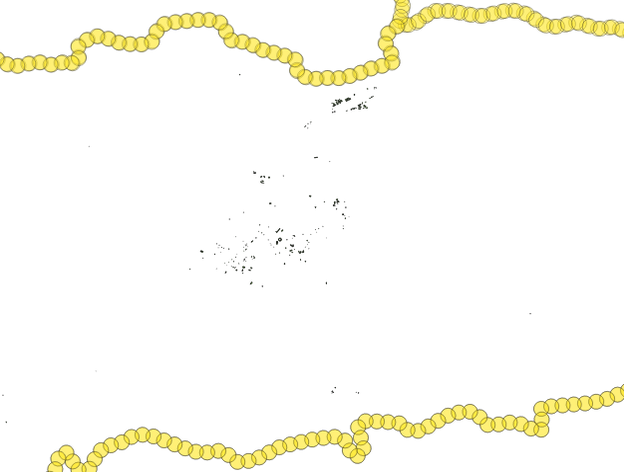} 
\\ \\
    & (a)Gwalior City Roads map &
      (b)Gwalior City Buildings map
  \end{tabular}
  \caption{The Gwalior city maps with with the Northern part Gwalior district within yellow contour}\label{figure2}
\end{figure}

\subsection{About Gwalior}
Gwalior district is located in the Northern Madhya Pradesh.  It comprises of four tehsils with Gwalior (formerly, Gird), Bhitarwar in the North and Dabra (formerly, Pichore), Chinour in the South. The district of Gwalior also comprises of four community development blocks viz.  Ghatigaon (Barai), Morar in the North and  Dabra, Bhitarwar in the South. The city of Gwalior is in the region connecting the two Northern blocks, ie. made of regions from the Ghatigaon and Morar blocks.

The population of Gwalior district as per 2011 census is 20,32,036. The population of Gwalior city is 10,69,276. 

Figure \ref{figure1} (a) and (b) show the Gwalior District and the Northern region of Gwalior district within yellow contour. 
Figure  \ref{figure1} (a), (b) ,(c) and (d)  show the Gwalior city details such as administrative map, population map, city roads map and city buildings map. The location map and city roads map have been overlaid on satellite images.

\subsection{GIS map sources}
The GIS maps were obtained from various sources on the Internet Although the COMOKIT model required only the buildings shape file. The buildings shape file is obtained from the  Openstreetmap dataset\cite{OpenStreetMap}. The population tiff files were obtained from Humanitarian data exchange \cite{hum}. The administrative areas shape files were downloaded from the global administrative dataset \cite{gadm}.

\subsection{Models Used}

In India, various measures such as school closures, complete lockdown and home containment were implemented at various instances to control the spread of the virus across the large population. Same measures were also implemented in the Gwalior region. This experiment compares the three measures such as school closure, home containment and no containment for the spatial map of Gwalior versus the number of deaths. 

Further,  the following containment actions were compared with No containment, Realistic lockdown with 10\% of essential workers and 20 daily tests, Family containment when positive member and  Dynamic spatial lockdown versus the number of deaths.
The dynamic spatial lockdown strategy factors in the overall deaths in a geographical region to take decisions on various containment measures to minimize the destruction in an uncertain environment. 

In the third set of experiments for containment policy decision making, the effect of partial testing, mass testing, no testing and no policy were used for comparision. The partial testing scenario is more realistic due to the lack of resources and lack of willingness in people to participate in mass testing. 

\subsection{Workflow} 
The preparation and execution workflow performed within the COMOKIT environment are as follows:
\begin{enumerate}
    \item Preparation of data 
    \item Setting program environment
    \item Models
\end{enumerate}
\subsubsection{Preparation of data}
The tutorial on using the COMOKIT to create a case study in \cite{cktut} was followed by performing the following steps.
\begin{enumerate}
\item Import the study area into GAMA softare.
\item Generate the build environment and satellite image
\item Retrieve the GoogleMap data with Qgis.
\item Use the GoogleMap data to assign types to buildings
\end{enumerate}
\subsubsection{Setting up program environment}
The epidemiological parameters of the COMOKIT model are setup through the parameter variables in Mode/Parameters.gaml. The paramters are listed under various heads such as Epidemiological parameters, Population parameters, Synthetic agenda parameters and Building type parameters. The epidemiological parameters given are further grouped under a. Environmental contamination dynamics b. Human to human transmission dynamics c. Hospitalization and severity d. Testing and mask wearing. The synthetic agent population generation dynamics is set through population parameters. The behaviour of the synthetic agents over time flow is caliberated using synthetic agenda generator and activity file. For the experiments conducted, the default parameters were used.
\subsubsection{Models}
The following gaml files under policy comparision component in the comokit modeling kit were executed with the Gwalior spatial map.
\begin{enumerate}
\item Comparison of three measures
\item Comparison of containments
\item Comparison of realistic actions
\end{enumerate}

\section{Results and Analysis} 
The software used for the experimentation was the Gama spatial agent modeling tool and the comokit covid 19 modeling toolkit. The experiments were run for around 5500 cycles.

\begin{figure}
\includegraphics[width=\textwidth]{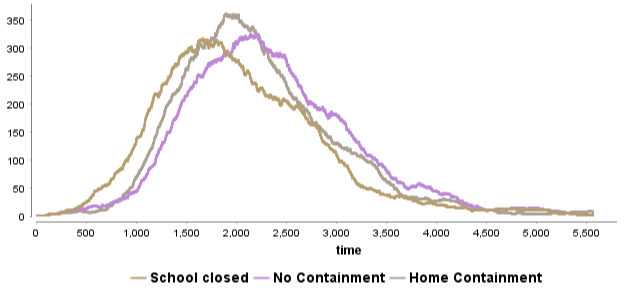}
\caption{A Comparison of the three measures} \label{fig7}
\end{figure}

Figure \ref{fig7} shows that all the three containment measures do not differ much in their effect in the long run with more or less similar peaks for the Gwalior .

\begin{figure}
\includegraphics[width=\textwidth]{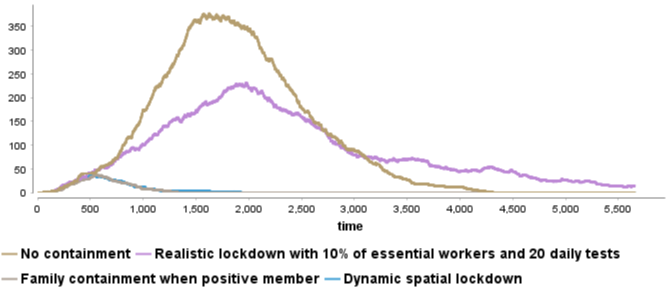}
\caption{A comparison of the containments} \label{fig8}
\end{figure}

Figure \ref{fig8} compares four different strategies for reducing disease spread. This figure shows that distinct differences between different containment strategies. The no containment strategy shows an immediate peak of the disease in the region. Realistic lockdown strategy is relatively better than the no- containment strategy. The family containment and dynamic spatial lockdown strategies if fully effected show the best possible response for the spatial map of Gwalior.

\begin{figure}
\includegraphics[width=\textwidth]{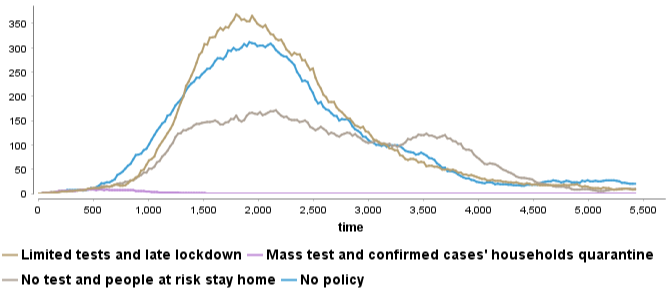}
\caption{A comparison of realistic actions} \label{fig9}
\end{figure}

Figure \ref{fig9} compares four realistic actions which include tests. Among the four, Limited tests with late lockdown is only as good as having no policy for the region. The mass testing and household quarantine of confirmed cases is the best strategy but not practical. Not performing any testing but ensuring that the risky people stay at home is still a better strategy than no policy and late lockdown policies.

From figures \ref{fig7}, \ref{fig8} and \ref{fig9}, it can be inferred that while mass testing really proves beneficial, dynamic spatial lockdown and sincere family containment on a member becoming positive are also better measures for containing the disease spread.

\section{Conclusion}
This paper presented some of the results obtained by using spatial data of Gwalior, India as input to the COMOKIT spatial modelling kit. The experiments to make policy decisions were executed and the results have been presented. On comparision, the results show that dynamic spatial containment strategy works well. This work can be further enhanced by altering the parameters used in the simulation framework to match with the realworld. A comparision of the obtained results with the realworld variations may further help in investigating identifying any other factors that may have contributed to the disease spread as it happened.

\section{Acknowledgement}
The author acknowledges DST-SERB for funding the work reported in this paper.

%
%
%
\bibliographystyle{splncs04}
\bibliography{ref}
%




\end{document}